\documentclass[10pt,letterpaper]{article}
\usepackage[top=0.85in,left=2.75in,footskip=0.75in]{geometry}

\usepackage{changepage}
\usepackage{color}
\usepackage[utf8]{inputenc}

\usepackage{textcomp,marvosym}

\usepackage{fixltx2e}

\usepackage{amsmath,amssymb}

\usepackage{cite}

\usepackage{nameref,hyperref}

\usepackage[right]{lineno}

\usepackage{microtype}
\DisableLigatures[f]{encoding = *, family = * }

\usepackage{rotating}

\raggedright
\usepackage[aboveskip=1pt,labelfont=bf,labelsep=period,justification=raggedright,singlelinecheck=off]{caption}

\makeatletter
\renewcommand{\@biblabel}[1]{\quad#1.}
\makeatother

\date{}

\usepackage{lastpage,fancyhdr,graphicx}
\usepackage{epstopdf}
\fancyheadoffset[L]{2.25in}
\fancyfootoffset[L]{2.25in}
\begin{document}
\vspace*{0.35in}

\begin{flushleft}
{\Large
\textbf\newline{Modeling Radicalization Phenomena in \\Heterogeneous Populations}
}
\newline
\\
Serge Galam\textsuperscript{1,\Yinyang},
Marco Alberto Javarone\textsuperscript{2,\Yinyang}\bigskip
\\
\bf{1} CEVIPOF – Centre for Political Research, CNRS and Sciences Po, Paris, France
\\
\bf{2} Dept. Mathematics and Computer Science, University of Cagliari, Cagliari Italy
\\
\bigskip

%
%
\Yinyang These authors contributed equally to this work.





* marcojavarone@gmail.com

\end{flushleft}

\section*{Abstract}
The phenomenon of radicalization is investigated within a mixed population composed of core and sensitive subpopulations. The latest includes first to third generation immigrants. Respective ways of life may be partially incompatible.
In case of a conflict core agents behave as inflexible about the issue. In contrast, sensitive agents can decide either to live peacefully adjusting their way of life to the core one, or to oppose it with eventually joining violent activities. 
The interplay dynamics between peaceful and opponent sensitive agents is driven by pairwise interactions. These interactions occur both within the sensitive population and by mixing with core agents. The update process is monitored using a Lotka-Volterra-like Ordinary Differential Equation.
Given an initial tiny minority of opponents that coexist with both inflexible and peaceful agents, we investigate implications on the emergence of radicalization. Opponents try to turn peaceful agents to opponents driving radicalization. However, inflexible core agents may step in to bring back opponents to a peaceful choice thus weakening the phenomenon.
The required minimum individual core involvement to actually curb radicalization is calculated.It is found to be a function of both the majority or minority status of the sensitive subpopulation with respect to the core subpopulation and the degree of activeness of opponents.
The results highlight the instrumental role core agents can  have to hinder radicalization within the sensitive subpopulation. Some hints are outlined to favor novel public policies towards social integration.


\section*{Introduction}
The phenomenon of radicalization~\cite{radicalization01} is of central interest in the context of criminality and terrorism. It is currently spreading all over the world including European countries. The recent unprecedented  terrorists attacks in Paris (November 13, 2015) and Brussels (March 22, 2016) took life of respectively 130 and 32 persons with over 300 wounded in each case ~\cite{lemonde01,ncb01}. It puts at a very high level the burden on making substantial progress in the mastering of the issue.
Over the years sociologists and social-psychologists have contributed a good deal of work to the phenomenon~\cite{radicalization01, radicalization02, radicalization03, radicalization04}. 
However an understanding, which could lead to some practical curbing of radicalization is unfortunately still lacking as dramatically demonstrated by the recent series of terrorist attacks in France (2015~\cite{lemonde02,lemonde01}) and in Brussels (2016~\cite{ncb01}). 
One promising direction is the prospect to access the huge amount of data (Big Data) which exists in the World Wide Web. It could open a valuable source of surveillance and forecasting to prevent some aspects of radicalization spreading. However, efficient data-mining tools are still to be constructed yet within the constraints related to the preservation of individual privacies. Accordingly, under the current risk of loosing control of the situation any new attempt to tackle the issue of radicalization is valuable in itself. To identify some hints to implement novel adequate policies towards at least the hindrance of radicalization spreading is of particular importance.
Along this line it happens that the modern field of sociophysics~\cite{galam01, loreto01, buechel01} where models inspired from physics are developed to describe a large spectrum of social behaviors, may contribute to the challenge. Among others, sociophysics includes the study of opinion dynamics~\cite{sznajd01,javarone01,javarone02,redner01}, language dynamics~\cite{loreto01,javarone04}, crowd behavior~\cite{loreto01}, criminal activities~\cite{perc05,galam03-bis,galam03-ter,javarone03}, and cultural dynamics~\cite{moreno02,iglesias01}.
Our work, focusing on a formal modeling of radicalization (see also~\cite{mcmillon01}) from the viewpoint of opinion dynamics, subscribes to this trend~\cite{galam05}. Therefore, according to the analytical approaches developed in sociophysics the proposed model adopts some assumptions that allow to simplify the scenario of reference. The complexity underlying terrorism phenomena is thus reduced to a series of more simple local interactions monitored by two parameters, which tune the global dynamics of the system. The focus on local interactions to reach the global equilibrium state constitutes one major trend of statistical physics, i.e., the branch of physics from which sociophysics developed.
More specifically we consider a mixed population made up of two subpopulations, each one sharing a peculiar way of life. First one is a core population locally rooted in the country. In contrast,  the other one is an immigrant (two, three generations) subpopulation whose way of life is rooted in another territory. Differences between the two ways of life may be strong, numerous and contradictory. 
However, in case of a disagreement about some specific cultural habit like for instance wearing the Islamic veil, both subpopulations do not stand at the same level of resilience. Core agents consider that it is not up to them to modify their way of life or accept from newcomers behaviors perceived as contrary to their long time country rooted cultural habits.
Core agents behave here as inflexible agents. For them it is up to newcomers including  immigrants even at second or third generation to adjust to the country prevailing way of life. It is thus up to newcomers to either choose to live peacefully with the core population adjusting part of their habits to the local constraints or to maintain the integrality of their habits at a cost of creating conflicts with the core population.
Accordingly, the newcomers can be considered as sensitive agents. They can choose between two individual states either peaceful or opponent. Sensitive agents are entitled to shift state from peaceful to opponent and vice-versa. In addition we make the assumption that being in an opponent state may lead the corresponding agent to take part or to support violent activities. In principle, the latter choice can be linked to the appearance of local terrorist groups.
We are dealing with a mixture of inflexible and sensitive agents in given fixed proportions $\sigma_I$ and $\sigma_S$ with $\sigma_I +\sigma_S=1$. However, $\sigma_S$ is made up of two time dependent parts $\sigma_P(t)$ and $\sigma_O(t)$, which are the respective proportions of peaceful and opponent sensitive agents. At any time $t$  $\sigma_S =\sigma_P(t) +\sigma_O(t)$.
The time dependence is driven by an internal dynamics among sensitive agents. It is the result of pairwise interactions both among themselves between peaceful and opponent agents and with inflexible agents. An opponent may drive a peaceful agent to opponent and an inflexible may drive peaceful an opponent agent. The associated dynamics is studied using a Lotka-Volterra-like Ordinary Differential Equation.
Given an initial tiny minority of opponents we investigate the role of their activeness~\cite{galam04} in turning peaceful agents to opponents via pairwise interactions. The effectiveness of their activism is materialized in the degree of radicalization of the sensitive population against the core population. It creates a social basis for passive supporters~\cite{galam03} to emerge in support to terrorists~\cite{islamic01}.
In parallel, the mechanism behind the dynamics of  radicalization enlightens by symmetry a potential role core inflexible agents could have in the launching of an eventual counter radicalization. By individual counter activeness core agents can contribute substantially to both curb the radicalization spreading and in certain conditions make it shrink down to an equilibrium state where inflexible, peaceful and opponent agents co-exist. The associated required minimum core individual involvement is calculated. It is found to be a function of both the majority or minority status of the sensitive subpopulation with respect to the core subpopulation and the degree of activeness of opponents. 
It is worth to stress that different mathematical frameworks could be used to describe our dynamics. For instance, approaches based on evolutionary game theory~\cite{holme01,perc03,nowak01,tomassini01} allow to perform both computational and analytical (e.g.,~\cite{javarone05}) investigations. It requires to define a payoff matrix and rules for local interactions to monitor the updating. 
In this work we use stochastic processes based on opinion dynamics~\cite{galam01}. Local interactions reduce to contact processes, which make updating rules to depend on the relative densities of the various agent states. The choice of the current approach in the modeling arises from the aim to evaluate to which extent the heterogeneity of a population in cultural and behavioral terms may lead to critical and complex social phenomena as radicalization. Furthermore, it is important to emphasize that the attribute `inflexible' adopted to describe the core population stand, refers to cultural habits and traditions which allow to peacefully coexist with individuals coming from abroad provided they share the fundamental features of the local cultural frame.
It happens that opinion dynamics constitutes one of the most investigated topics in sociophysics and in computational social science. 
For instance, its dynamics have been recently studied using the framework of multiplex networks~\cite{battiston01,li01} considering different social behaviors~\cite{krokidakis01,krokidakis02,krokidakis03,pickering01,cheon01}. It allows to understand phenomena recorded in huge social network datasets~\cite{menezes01,menezes02,girvan01}. 
Opinion dynamics allows to analyze and to model the spreading of ideas, opinions, and feelings by reducing the study of complex social scenarios to the analysis of few variables~\cite{xie01}. Even terrorism and criminal activities may be studied by the same approach, i.e., reducing the related process to a problem of opinion dynamics.
To conclude, our results may contribute to shed a new light on the instrumental role core agents could play to curb radicalization and establish a coexistence with the sensitive population. Some hints at novel public policies towards social integration are obtained.
\section*{Previous Models}\label{sec:model_summary}
In the last years several authors have worked on opinion dynamics models to analyze various underlying behaviors, which produce social phenomena, e.g., group polarization, conformity and extremism. In this section, we briefly review some of these investigations, which are connected to our work along the topic of extreme social phenomena, especially radicalization.
A computational model for tackling political party competitions is introduced in~\cite{diaz01}. The authors investigate different possible occurrences of fragmentation according to variations in the amount of important political issues and their current relevance. Different interaction patterns among voters are considered using an analytical approach. The focus is on the role of extremism in opinion dynamics with a qualitative analysis of real scenarios.
The complex social phenomenon of group polarization is described in~\cite{dixit01} in the context of politics. In particular, the authors propose a model based on probability theory to drive the emergence of group polarization. The emergence of risks is shown to be related to the group polarization in a wide range of scenarios related to political and economical issues (e.g., immigration, religion, welfare state, human rights). The results highlight the necessity to a better understanding of the emergence of extreme opinions.
The connection between contradictory public opinions, heterogeneous beliefs and the emergence of extremism is analyzed in~\cite{galam06}. An agent-based model considers a population with different socio-cultural classes to describe the process of opinion spreading with calculations performed on small groups of individuals (e.g., composed of $3$ and $4$ agents). The model constitutes a useful reference for defining models related to complex social phenomena. Moreover, the related results suggest that the direction of the inherent polarization effect, which occurs in the formation of a public opinion driven by a democratic debate, is biased due to the existence of common beliefs within a population. 
Opinion dynamics is also studied using computational approaches, e.g., by agent-based models on continuous or discrete spaces. Such approaches require a careful attention during the implementation phase. For instance, in the work~\cite{cioffi01} authors focus on the role of activation regimes. More precisely they compare different asynchronous updating schemes (e.g., random and uniform). The activation regime refers to the order or scheme adopted to let agents express their opinion. As a result, the activation regime is found to affect opinion dynamics processes in some cases (i.e.,~\cite{cioffi01}). It is therefore of importance to clearly state which activation regime is selected to implement a dynamical model.
The role of conformity in the q-voter model by arranging agents on heterogeneous networks has been also investigated ~\cite{javarone06}. The authors showed that different steady states may be reached by tuning the ratio of conformists versus that of nonconformists in an agent population, which evolves according to the dynamics of the q-voter model. In our model we do not consider complex topologies. However, the influence that may arise from different interaction patterns may constitute the topic of future investigations.
%
%
\section*{Mathematical Model}\label{sec:model}
In order to study the emergence of radicalization in an heterogeneous population we consider a system with $N$ interacting agents distributed among inflexible ($I$), peaceful ($P$) and opponent ($O$) agents.
Each category refers to a different behavior or feeling. Inflexible and opponent agents have behaviors mapped respectively to states $s = \pm 1$. Peaceful agents have a behavior mapped to the state $s = 0$.
Inflexible agents never change state (see also~\cite{galam02}) while peaceful and opponent agents may shift state from one to another over time. Opponents may become peaceful and peaceful may become opponents. Hence, neither peaceful nor opponent agents may assume the state of inflexible agents.
Inflexible agents interact with sensitive agents both peaceful and opponents. During these pairwise interactions when an inflexible agent meets an opponent it may well turn the opponent to peaceful via different paths. Among those paths most are spontaneous through normal social and friendship practices. But as it will appear latter, exchanges could become intentional as to promote coexistence with sensitive agents via monitored informal exchanges. To account for all interacting pairs a parameter $\alpha$ is introduced to represent on average the rate per unit of time of encounters where opponents become peaceful agents.
In parallel and in contrast we introduce the parameter $\beta$ to account on average for the rate of success of opponents in convincing peaceful agents to turn opponents. Contrary to inflexible agents opponents are acting intentionally to increase the support to their radical view within the sensitive population. The value of $\beta$ is a function of the power of conviction of opponents. It also takes into account the activeness of opponent agents since opponents are activists. It is not the case of the core inflexible agents who interact spontaneously with sensitive agents without an a priori goal. It is worth to stress that both $\alpha$ and $\beta$ may in principle vary over time.
However, the corresponding time scale for variation is expected to be much longer than the time scale of the dynamics driven by pairwise interactions. This is why at the present stage of our work $\alpha$ and $\beta$ are assumed to be fixed and constant. Analyzing their time dependence, which might be of great interest to get further insights on equilibriums among people belonging to different cultures is left for future work.
We emphasize that our analytical approach entails to consider the system as if it was continuous, i.e., analyzing the relative densities of agents in the various states. A similar approach is usually followed in other contexts as epidemic dynamics~\cite{vespignani01, lagorio01}.
A compartmental approach to the studying of epidemics entails to analyze the spreading of a disease by modifying the state of agents. For instance, the SIS~\cite{epidem01} model considers a two-state population where agents may vary their state from $S$ (i.e., susceptible) to $I$ (i.e., infected) and vice-versa over time. Considering the probability to get infected or to heal the dynamics can be studied analytically defining ODEs as if the underlying system were continuous.
Going to the analytical details of our model we defined the following system of equations
\begin{equation}\label{eq:evolution}
\begin{cases}
\frac{d\sigma_P(t)}{dt} = \alpha \sigma_I \sigma_O(t) - \beta \sigma_O(t) \sigma_P(t)\\ 
\frac{d\sigma_O(t)}{dt} =  \beta \sigma_O(t) \sigma_P(t) - \alpha \sigma_I \sigma_O(t)\\
\sigma_I + \sigma_P(t) + \sigma_O(t) = 1
\end{cases}
\end{equation}
\noindent where $\sigma_I$ is the constant density of inflexible agents, while $\sigma_O(t)$ and $\sigma_P(t)$ are the respective densities of peaceful and opponent agents at time $t$.
Dealing with densities the third equation of system~\ref{eq:evolution} allows to reduce the number of ODEs to one equation. In particular, choosing the peaceful agents density $\sigma_P(t)$ we get
\begin{equation}\label{eq:evolution_reduced}
\frac{d\sigma_P(t)}{dt} =  \alpha \sigma_I (1 - \sigma_I - \sigma_P(t)) - \beta (1 - \sigma_I - \sigma_P(t)) \sigma_P(t)
\end{equation}
The equilibrium state of the population can be obtained from $\frac{d\sigma_P(t)}{dt} = 0$, which reads
\begin{equation}\label{eq:evolution_equilibrium}
\beta  \sigma_P(t)^2-  ( \alpha \sigma_I + \beta(1 - \sigma_I))\sigma_P(t)+\alpha \sigma_I  (1 - \sigma_I))=0
\end{equation}
The two solutions of equation~\ref{eq:evolution_equilibrium} read
\begin{equation}\label{eq:equilibrium_solution}
<\sigma_P> = \frac{\alpha \sigma_I + \beta(1 - \sigma_I) \pm \sqrt {[\alpha \sigma_I + \beta (1 - \sigma_I)]^{2} - 4 \beta \alpha \sigma_I (1 - \sigma_I)}}{2 \beta}
\end{equation}
\noindent where $<\sigma_P>$ is the equilibrium value of peaceful agents. Those values simplify to
\begin{equation}\label{eq:equilibrium_solution_value}
<\sigma_P> = \begin{cases} 
1 - \sigma_I\equiv p_1\\ 
\frac{\alpha}{\beta} \sigma_I\equiv p_2
\end{cases}
\end{equation}
which implies the two associated equilibrium opponent values 
\begin{equation}\label{eq:equilibrium_solution_value_op}
<\sigma_O> = \begin{cases} 
0\\ 
1 -  \frac{\alpha+\beta}{\beta} \sigma_I
\end{cases}
\end{equation}
Indeed equation~\ref{eq:evolution_reduced} can be solved analytically to yield
\begin{equation}\label{eq:solution_neutral}
\sigma_P(t) = p_2+\frac{p_1-p_2}{1 - \frac{\sigma_P(0)-p_1}{\sigma_P(0)-p_2}  e^{\beta(p_1-p_2) t}}
\end{equation}
Fig 1 shows the evolution of the system on varying the initial conditions.
\begin{figure*}
\centering
\includegraphics[width=1.0\textwidth,natwidth=610,natheight=642]{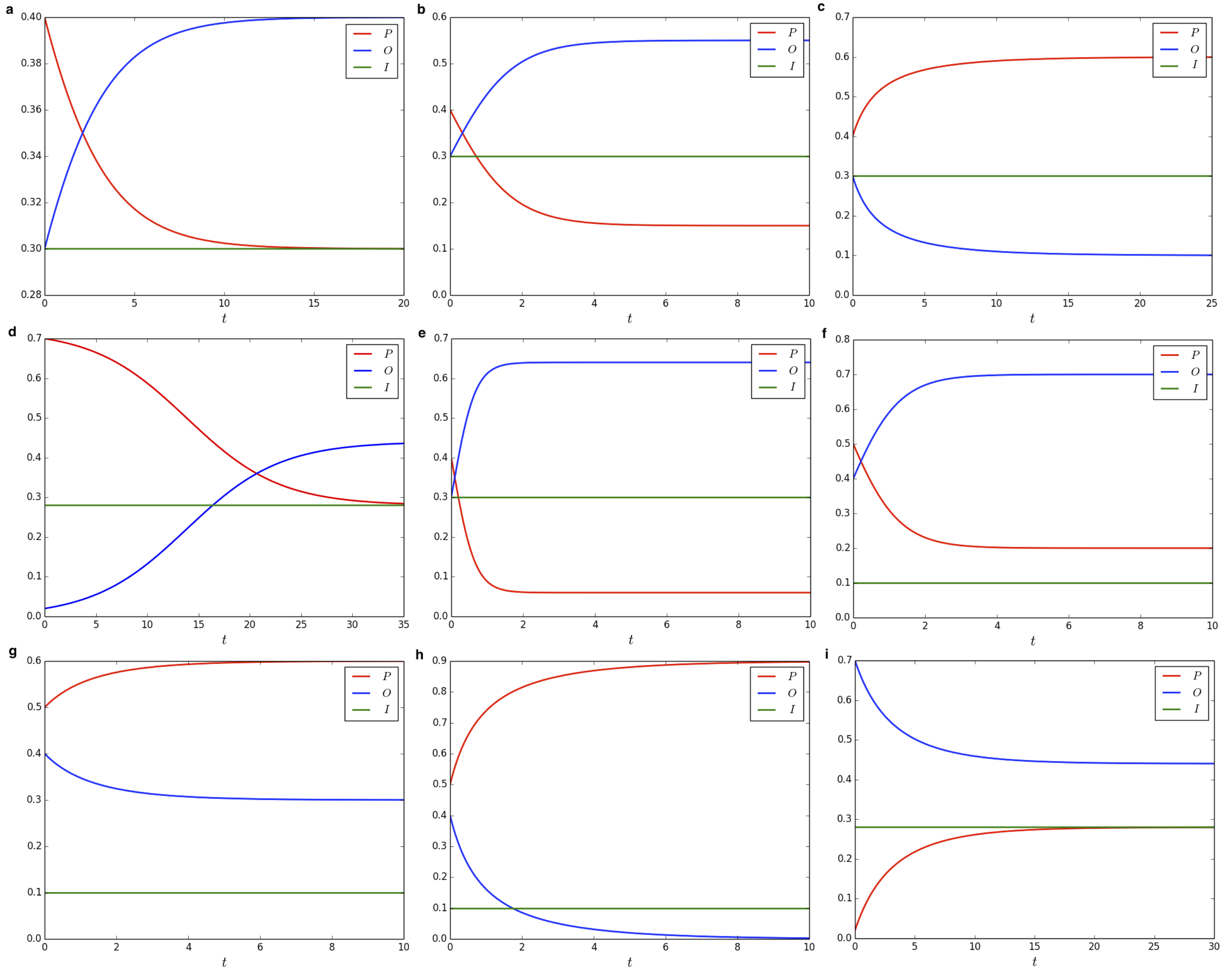}
\caption{Evolution of the system on varying initial conditions: \textbf{a} $\sigma_I= 0.3$, and $\sigma_O = 0.3$, $\alpha = 1.0$, $\beta = 1.0$. \textbf{b} $\sigma_I = 0.3$, and $\sigma_O = 0.3$, $\alpha = 1.0$, $\beta = 2.0$. \textbf{c} $\sigma_I = 0.3$, and $\sigma_O = 0.3$, $\alpha = 4.0$, $\beta = 2.0$. \textbf{d} $\sigma_I = 0.28$, and $\sigma_O = 0.02$, $\alpha = 0.5$, $\beta = 0.5$. \textbf{e} $\sigma_I = 0.3$, and $\sigma_O = 0.3$, $\alpha = 1.0$, $\beta = 5.0$. \textbf{f} $\sigma_I = 0.1$, and $\sigma_O = 0.4$, $\alpha = 4.0$, $\beta = 2.0$. \textbf{g} $\sigma_I = 0.1$, and $\sigma_O = 0.4$, $\alpha = 12.0$, $\beta = 2.0$. \textbf{h} $\sigma_I = 0.1$, and $\sigma_O = 0.4$, $\alpha = 22.0$, $\beta = 2.0$. \textbf{i} $\sigma_I = 0.28$, and $\sigma_O = 0.7$, $\alpha = 0.5$, $\beta = 0.5$.}\label{fig:evolution}
\end{figure*}
\subsection{Analysis of the Stability}
We analyze the respective stability ranges for $p_1$ and $p_2$:
\begin{equation}\label{eq:stability}
\frac{d\sigma_P}{dt} (\sigma_P) \simeq \frac{d\sigma_P}{dt}(<\sigma_P>) + (\sigma_P - <\sigma_P>) \lambda
\end{equation}
\noindent where $\frac{d\sigma_P}{dt}(<\sigma_P>) =0$ and $\lambda \equiv \frac{d^2\sigma_P}{dt d \sigma_P}|_{<\sigma_P>}$, we obtain
\begin{equation}\label{eq:lambda_value}
\lambda = -[\alpha \sigma_I + \beta(1 - \sigma_I)] + 2\beta \sigma_P
\end{equation}
\noindent Therefore, for respective values $p_1,p_2$ we obtain
\begin{equation}\label{eq:lambda_value_solutions}
\begin{cases} 
\lambda_1=-\alpha \sigma_I + \beta (1 - \sigma_I)=\beta(p_1-p_2)\\ 
\lambda_2=\alpha \sigma_I - \beta (1 - \sigma_I)=-\beta(p_1-p_2)
\end{cases}
\end{equation}
\noindent Stability being achieved for $\lambda < 0$, equation \ref{eq:lambda_value_solutions} shows that $p_1(p_2)$ is stable when $p_1<p_2(p_1>p_2)$. 
Accordingly we get two stable regimes:
\begin{equation}\label{eq:final_stability}
\begin{cases} 
p_1\le p_2\Leftrightarrow\sigma_I \ge I_c  \\ 
p_1\ge p_2\Leftrightarrow \sigma_I \le I_c  
\end{cases}
\end{equation}
\noindent with $I_c\equiv\frac{\beta}{\alpha + \beta}$. These two regimes yield the respective
equilibrium values for peaceful and opponent agents as from ~\ref{eq:equilibrium_solution_value} and ~\ref{eq:equilibrium_solution_value_op} 
\begin{equation}\label{eq:final_stability}
\begin{cases} 
 <\sigma_P> = p_1=1 - \sigma_I, <\sigma_O >= 0 \\ 
<\sigma_P> =p_2= \frac{\alpha}{\beta}\sigma_I, <\sigma_O> = 1-\frac{\sigma_I}{I_c}=p_1-p_2
\end{cases}
\end{equation}
The first equation of system~\ref{eq:final_stability} highlights that in some conditions the amount of opponent agents is equal to zero. Hence, we perform a further investigation to study under which conditions it is possible to avoid the phenomenon of radicalization (i.e., by reaching the equilibrium state $<\sigma_O >= 0$). In terms of opinion dynamics these results indicate that under appropriate conditions it is possible to remove one opinion from the system. Given the relevance of this outcome in the related context, i.e., criminal activities and terrorism, we explore in more details this result.
\subsection{Extinction processes}
From the above results radicalization can be totally thwarted if $\sigma_I \ge I_c$. 
Accordingly, given $\sigma_I$ and $\beta$ the individual involvement for the inflexible population in striking up with individual opponents must be at least at a level
\begin{equation}\label{eq:alpha}
\alpha > (\frac{1}{\sigma_I} - 1)\beta
\end{equation}
Therefore, as seen from equation \ref{eq:alpha} the larger $\sigma_I$ the less effort is required from the inflexible population.
However, the more active are the opponents (i.e., larger $\beta$) the more involvement is required. To visualize the multiplicative factor by which $\alpha$ must overpass $\beta$ it is worth to draw the curve $\frac{1}{\sigma_I}-1$ as a function of $\sigma_I$ as shown in Fig 2.
\begin{figure*}
\centering
\includegraphics[width=0.75\textwidth,natwidth=610,natheight=642]{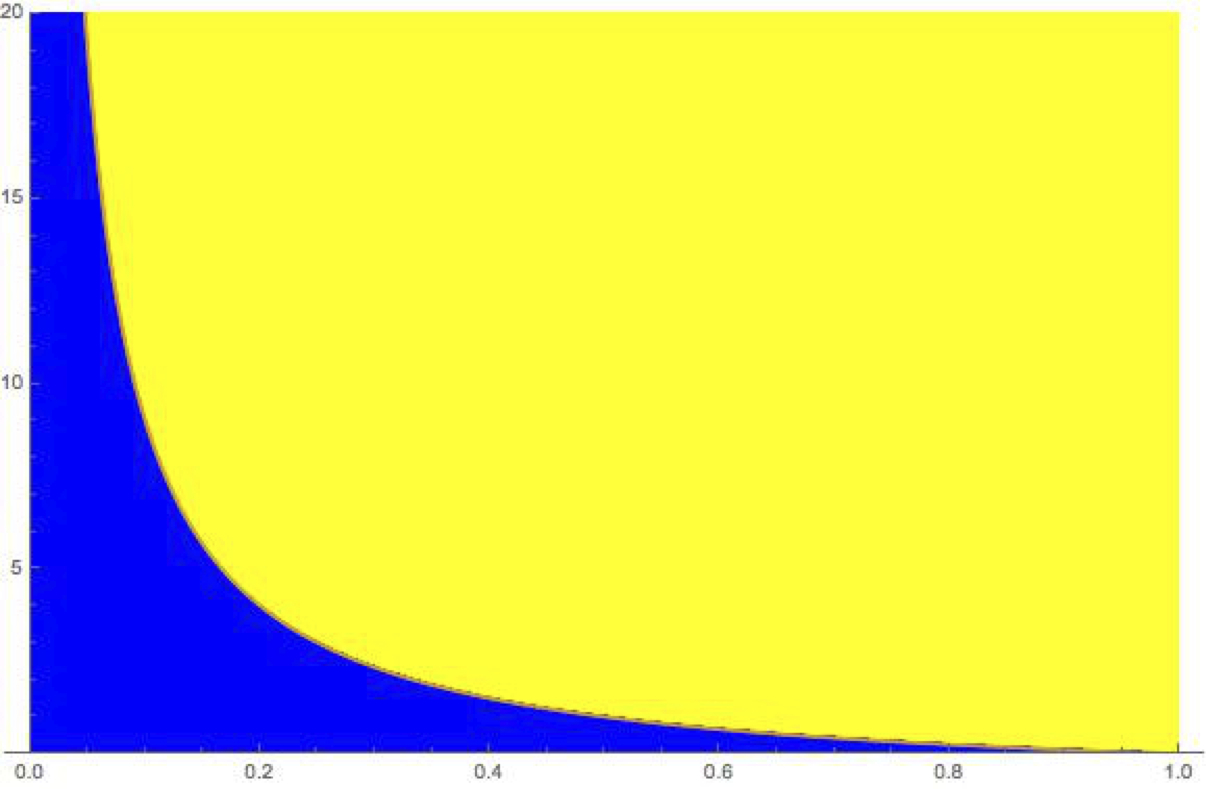}
\caption{The curve $\frac{1}{\sigma_N}-1$ is shown as a function of $\sigma_N$. All cases for which  the value of $\frac{\alpha}{\beta}$ is above the curve (yellow, clear)  correspond to situations for which radicalization is totally thwarted. When the value of $\frac{\alpha}{\beta}$ is below the curve (blue, dark) radicalization takes place on a permanent basis.
}\label{fig:a=0}
\end{figure*}
From Equation~\ref{eq:alpha} it is seen that to prevent radicalization inflexible agents's involvement must be either lower ($\alpha < \beta$) or larger ($\alpha > \beta$) than that of opponents depending on the magnitude of the multiplicative factor $\frac{1}{\sigma_I}-1$.
When $\frac{1}{\sigma_I}-1<1$, i.e., $\sigma_I>\frac{1}{2}$ core agents do not need to much individual engagement as could be expected in the case of a coexistence of a core majority population with a sensitive minority subpopulation. More precisely, the engagement depends on the opponent activism but the core population benefiting from its majority status. In this case its requirement is always lower than the opponent involvement. However, the situation turns difficult when the initial sensitive minority turns to a majority status as it occurred in some specific urban areas. In that case to avoid a radicalization requires a very high individual engagement from the core agents, which may be rather hard to implement. In particular since no collective information is available about the situation.
We thus have three different cases: \textbf{1)} $\sigma_I>\frac{1}{2}$, \textbf{2)} $\sigma_I=\frac{1}{2}$, and \textbf{3)} $\sigma_I<\frac{1}{2}$ to consider to determine the respective level of individual core involvement to avoid the phenomenon of radicalization.%
\\
\textbf{Case 1.} 
For $\sigma_I > \frac{1}{2}$ core agents need little involvement to thwart totally the radicalization of the sensitive subpopulation with values of $\alpha$ much lower than $\beta$. Indeed, opponent agents need to produce very high values of $\beta$ (compared to $\alpha$) to survive, precisely the condition $\beta \ge \frac{\alpha \sigma_I}{(1- \sigma_I)}$ must be satisfied. However, very large values of $\beta$ can shrink to zero the amount of peaceful agents yielding a fully radicalized sensitive population, which although in a small minority status may produce substantial violence against inflexible agents.\\
\textbf{Case 2.} 
For $\sigma_I = \frac{1}{2}$ the opponent activism must be counter with an equal core counter activism since $\alpha \ge \beta$ makes opponent agents to extinct. Instead, for $\alpha < \beta$ peaceful and opponent agents coexist and the former disappear for large values of  $\beta$ with again a fully radicalized sensitive population with $<\sigma_O>=  \sigma_I$.\\
\textbf{Case 3.} 
For values $\sigma_I<\frac{1}{2}$, if $\alpha = \beta$ the equilibrium condition entails that $<\sigma_P> = \sigma_I$ (and $<\sigma_O> = 1 - 2\sigma_I$).
If $\alpha > \beta$, we can reach the extinction of opponent agents as $\frac{\sigma_I}{I_c} = 1$. 
In contrast when $\alpha < \beta$ opponent agents strongly prevail in the population.\\
\subsection{Degree of radicalization}
In order to asses the degree of radicalization in a population we can introduce two parameters: $\zeta$ and $\eta$.
The former is defined to evaluate the fraction of opponent agents among flexible agents while the latter (i.e., $\eta$) evaluates the ratio between opponent and inflexible agents.
Therefore, $\zeta$ represents the relative ratio of opponents among flexible agents and $\eta$ gives a measure about the real power of opponents agents in a population. 
An high value of $\zeta$ (i.e., close to $1$) in a population with $\sigma_I >> 0.5$ indicates that strategies to fight radicalization are too weak but at the same time opponents are few. Therefore, in this case governments should take an action even if the situation seems still under control.
On the other hand, a low value of $\zeta$ (i.e., close to 0) together with a high value of $\eta$ represent an alarming situation. Indeed, even if there are only a few opponents among flexible agents their amount is bigger than that of inflexible ones~\cite{coles01}.
To evaluate these measures, $\zeta$ and $\eta$ have been defined as follows
\begin{equation}\label{eq:radicalization}
\begin{cases} 
\zeta = \frac{\sigma_O}{1 - \sigma_I}\\ 
\eta = \frac{\sigma_O}{\sigma_I}
\end{cases}
\end{equation}
\noindent hence, recalling that $\sigma_O = 1 - \sigma_I - \sigma_P$ and having solved analytically $\sigma_P(t)$ (see \ref{eq:solution_neutral}) we are able to compute values of both parameters $\zeta$ and $\eta$ at equilibrium and on varying the initial conditions ---see Fig 3.
\begin{figure*}
\centering
\includegraphics[width=1.0\textwidth,natwidth=610,natheight=642]{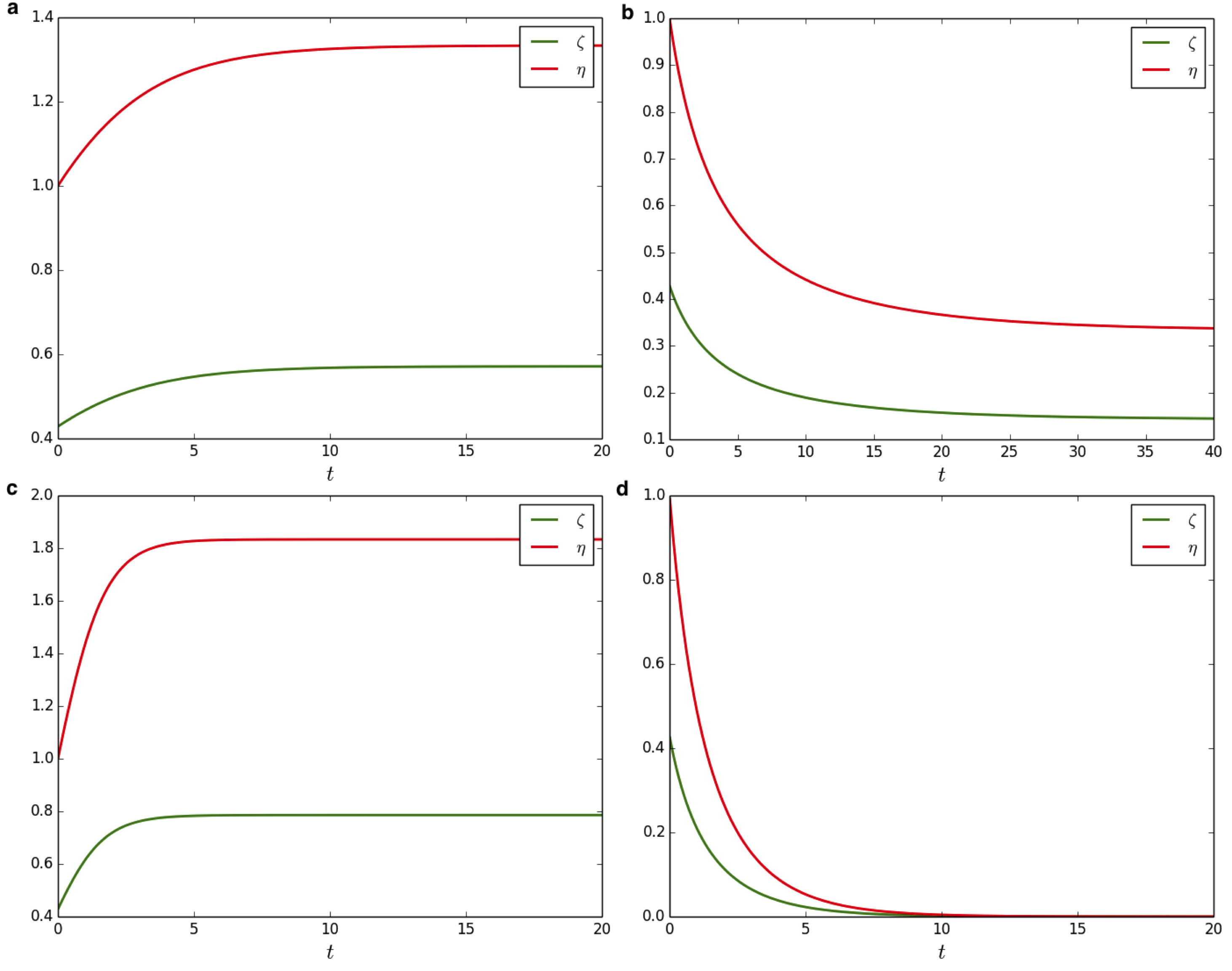}
\caption{\small Radicalization degree quantified according to the parameters $\zeta$ and $\eta$, on varying initial conditions: \textbf{a} $\sigma_I= 0.3$, and $\sigma_O = 0.3$, $\alpha = 1.0$, $\beta = 1.0$. \textbf{b} $\sigma_I = 0.3$, and $\sigma_O = 0.3$, $\alpha = 2.0$, $\beta = 1.0$. \textbf{c} $\sigma_I = 0.3$, and $\sigma_O = 0.3$, $\alpha = 1.0$, $\beta = 2.0$. \textbf{d} $\sigma_I = 0.3$, and $\sigma_O = 0.3$, $\alpha = 4.0$, $\beta = 1.0$.}\label{fig:radicalization}
\end{figure*}
It is worth to note that the parameter $\zeta$ as defined in~\ref{eq:radicalization} has a range in $[0,1]$. At equilibrium $\zeta = 0$ means that there are no opponent agents in the population while $\zeta = 1$ means that all flexible agents became opponents. 
On the other hand, the parameter $\eta$ has potentially an unlimited range from $0$ to $\infty$ (in the case $\sigma_I$ is very close to $0$ and $\sigma_O$ to $1$).
To conclude, we want to emphasize the meaningful role of the two parameters $\zeta$ and $\eta$. They represent a way to quantify in which extent radicalization phenomena are taking place in a population. Moreover, in more general terms we envision a further utilization of these parameters in opinion dynamics since they clearly indicate the prevalence of one opinion/state over another one.

\section*{Policy implications of the results}\label{sec:discussion}
The recent anti-western terrorist attacks~\cite{islamic01} in Europe have brought the question of radicalization at a top priority of policy maker agenda of the different European governments. In particular, most of the terrorists involved in the various killings which took place in several European capitals were either National citizens or legal residents. This very fact points to the direct link existing between terrorism and radicalization~\cite{aronson01}. 
Indeed, various institutions are faced with the difficult issue to implement innovative procedures to stop if not eradicate radicalization. The task turns out to be rather hard since radicalization has been prospering quietly in different areas of European countries for now many years without any substantial barrier. It has been a sensitive political issue and most officials had preferred the laissez-faire instead of addressing the problem in solid terms.
The dramatic scores of $2015$ Paris and $2016$ Brussels attacks have now prompted the necessity to face the problem and start implementing counter measures. The burden is on European governments to find ways to tackle the radicalization. Almost everyone is expecting action from the states. But the states seem to have no solid scheme to apply. One direction has been along the education side with the setting up of so-called de-radicalization programs. However, such an approach concerns identified radicals who have been arrested. All efforts and thoughts are focused on acting on radicalized citizens. Coercive measures are implemented against known associations and active leaders. Even to contain radicalization appears to be a challenging task.
Our model, although rather simple puts light on the process by which the phenomenon of radicalization spreads over within a sensitive population. It articulates around the capacity of radicalized agents to turn radical otherwise peaceful agents who had chosen to coexist with the native population sharing their habitat. This capacity is embedded in the coefficient $\beta$. In addition, the main novelty of our model is to account for the possible capacity of native agents to overturn radical agents in making them choose the peaceful state quantified with the coefficient $\alpha$.
Moreover, the ratio of native versus sensitive populations ($\frac{1-\sigma_I}{\sigma_I }$) was found to be a critical parameter.  In the past this ratio was rather stable over time with slow evolution. It made feasible to evaluate the activeness of radicals, which has been not meaningful for decades. However, rather quick changes may occur in the demography of the sensitive populations especially with the substantial increase of recent years immigration.
On this basis, our results show how the overall situation can be totally put upside down ($\frac{1-\sigma_I}{\sigma_I}<1 \rightarrow \frac{1-\sigma_I}{\sigma_I }>1 $) with respect to the extent of current radicalization while not much seems to have happened with respect to radical activities. Keeping the same level of activeness from radicals, a slow change in the population ratio may produce a sudden spreading of radicalization. Therefore the knowledge of the evolution of the current ratio of populations is a key parameter to evaluate the associated potential of radicalization spreading. And yet, in many countries like France, ethnic statistics are forbidden. 
Most of curbing radicalization still involves the state and diverse official institutions. In contrast the phenomenon of radicalization results from informal interactions among sensitive and radical agents. At this point our results unveiled a new and unexpected promising path to fight radicalization. 
An innovative strategy could be implemented by launching a citizen counter radicalization movement mapped from the path used by opponents to spread radicalization within the sensitive population. Instead of being the sole prerogative of National Authorities de-radicalization would become a citizen matter.
The same way a radical tries to turn a peaceful sensitive agent to hostility towards the natives, natives can try to bring back opponents to peaceful coexistence. Normal citizens would have to engage in personal interactions with sensitive agents to establish a solid ground for coexistence. The required degree of efficient citizen involvement can be clearly identified using the degree of activeness of the radicals and the ratio of subpopulations, native versus sensitive. While this ratio is at the exclusive hand of national authorities, the citizen involvement is a citizen prerogative.
In addition, the centrality of the ratio of subpopulations within a given territory emphasizes the importance of avoiding a de-mixing of the subpopulations. In case of a different discriminating distribution of the subpopulations within distinct sub-territories, radicalization would be enhance at once with the same proportion of radical due the large value of $\frac{1-\sigma_I}{\sigma_I }$ within the sub-territory where the sensitive population is mostly confined.

\subsection{Remark}
It is worth to notice that the proposed model may in principle be applied also to criminal and terrorist scenarios in homogeneous populations as it occurred in the cases of Italian Red Brigades and French Revolution. These two cases are concerned with homogeneous populations as both inflexible, peaceful and opponent agents belong to the local core population.
In the former case (i.e., red brigades) inflexible agents represent individuals who respect laws and believe in institutions and governments. Individuals having a different behavior can fall in the mild category of peaceful agents or in the extreme category of opponents (i.e., criminals).
Instead, in the case of the French revolution inflexible agents represent the small proportion of French nobility. The remaining part of the population is represented by peaceful and opponent agents. There, the extremely difficult life conditions fed opponent ideals and the wide proportion of the sensible subpopulation became completely opponent giving rise to a revolution.

\section*{Conclusion}\label{sec:conclusion}
To summarize we have identified the equilibrium state of a mixed population in terms of order or disorder phases. We have also identified the ratio between social strategies and the strength of opponents' ideal.
Since we refer to the concept of social strategies it is worth to emphasize that although the considered scenario can be modeled in various ways as those based on evolutionary games. There the concept of ``strategy'' acquires a particular meaning. Here we develop a model based on opinion dynamics processes. As a result social strategies are embodied in a parameter while updating rules depend on the density of different opinions in the population. Moreover, in this context opinions refer to the different cultural extractions and behaviors that can be observed in an heterogeneous population.
We remark that today the question of de-radicalization has became a key priority issue of internal security in European countries. Yet the challenge is intact with no ready to use solution.
Different state agencies are launching a series of experimental treatments but all are concerned with institutional managing of the issue. Given the acute current terrorist threat people are expecting and requesting policy makers to take initiative to curb the current phenomenon of radicalization within sensitive local populations.
Unlike this heavy policy trend our study has enlightened the crucial role so called ``normal citizens'' could play to stop the spreading of radicalism. It could even shrink it back with a serious perspective to eventually eradicate the actual growing threat set in European cities.
From our results it appears that an efficient action should not be limited to state involvement but also to call on individual voluntary engagement within their respective neighborhoods towards the sensitive individuals. 
Given an evaluation of radical activeness within some sensitive neighborhoods we were able to calculate the required degree of  ``normal citizen'' counter-activeness to curb radicalization. 
This degree of engagement was also found to depend on the ration of native to sensitive populations.
Focusing on local interactions in the modeling of these dynamics underlines the instrumental role neighborhood compositions can have in the shaping of the social behavior of the corresponding subpopulations. Today it often happens that people of different cultures are fostered to coexist together in the same district of a city occupying each a series of connected blocks.
The connection with natives is thus drastically reduced jeopardizing opportunities of real integration even after a few generations. In addition, within some delimited urban areas the majority group is no longer the native subpopulation. At the same time it is still majority in a close by other area. Accordingly given the same tiny proportion of opponents with the same degree of activeness in the two neighborhoods, one ends up highly radicalized while the other stays very peaceful. Local interactions and the degree of mixing are key factors to undermine the spreading of radicalization.
A free and uncontrolled (by authorities) settling of people often leads to a geographical concentration of sensitive subpopulations. As a result this process may spontaneously develop a natural ground for the emergence of hate towards native individuals. People are thus lead towards the strengthening of the initial culture differences, which results in the establishment of social distances with native individuals despite being physically very close to them.
To conclude, we want to highlight that our work creates a first step to envision new policies to support campaigns promoting the daily life sharing among people from different cultural backgrounds. In particular, we focus on methods that potentially may lead ``radical neighbors'' to the choice of coexistence, i.e., renouncing to fight against the native population. At least we hope our results will trigger more research along this path of individuals engaging to establish a peaceful coexistence with sensitive agents.
At this stage further studies along this direction are required, in particular from a computational social science perspective. It should be possible to identify earlier traces (i.e., Big Data) and seeds of radical behavior in social networks. Suitable tools to quantify their strength are also required.
Last but not least, we would like to stress that although we have been mentioning criminal activities we are not judging neither the motivations nor the ideal of opponent agents. Indeed, they can be considered negative (as in the case of current anti-western terrorism) or positive (as today in the case of the French revolution) depending both on the side taken and the chosen epoch. Our aim was to study the conditions of emergence or vanishing of radicalization as a social phenomenon independently of a moral judgment.

\section*{Acknowledgments}
MAJ would like to thank Fondazione Banco di Sardegna for supporting his work. This work was supported in part by a convention DGA-2012 60 0013 00470 75 01.


\begin{thebibliography}{10}
%

\bibitem{radicalization01}
Borum R. 
\newblock {{R}adicalization into Violent Extremism I: A Review of Social Science Theories.}
\newblock Journal of Strategic Security. 2011;4(4): 7--36

\bibitem{lemonde01}
\newblock{Milko, Marie, Salah, Elodie… les victimes des attentats du 13 novembre.}
\newblock Le Monde. 2015; Available: \url{http://www.lemonde.fr/attaques-a-paris/article/2015/11/15/guillaume-quentin-marie-les-victimes-des-attentats-du-13-novembre_4810428_4809495.html}.

\bibitem{ncb01}
\newblock{Brussels Attacks}
\newblock NBC News. 2016; Available: \url{http://http://www.nbcnews.com/storyline/brussels-attacks}.

\bibitem{radicalization02}
Thompson RL.
\newblock{Radicalization and the Use of Social Media.}
\newblock Journal of Strategic Security 2011; 4(4):167--190

\bibitem{radicalization03}
Haines HH.
\newblock{Black Radicalization and the Funding of Civil Rights.}
\newblock Social Problems. 1984; (32)1: 31--43

\bibitem{radicalization04}
Kruglanski AW, Gelfand MJ, Belanger JJ, Sheveland A, Hetiarachchi M,  Gunaratna R.
\newblock{The Psychology of Radicalization and Deradicalization: How Significance Quest Impacts Violent Extremism.}
\newblock Advances in Political Psychology. 2014; 35(1).

\bibitem{lemonde02}
\newblock{Charlie Hebdo. visé par une attaque terroriste, deuil national décrété.}
\newblock Le Monde. 2015; Available: \url{http://www.lemonde.fr/societe/article/2015/01/07/attaque-au-siege-de-charlie-hebdo_4550630_3224.html#}.

\bibitem{galam01}
Galam S.
\newblock {Sociophysics: a review of Galam models.}
\newblock International Journal of Modern Physics C. 2008; 19(3):409--440.

\bibitem{loreto01}
Castellano C, Fortunato S, Loreto V.
\newblock {Statistical physics of social dynamics.}
\newblock Rev. Mod. Phys. 2009; 81(2): 591--646.

\bibitem{buechel01}
Buechel B, Hellmann T, Klobner S.
\newblock{Opinion dynamics and wisdom under conformity.}
\newblock Journal of Economic Dynamics and Control 2015; 52: 240--257.

\bibitem{sznajd01}
Sznajd-Weron K, Sznajd J.
\newblock{Opinion Evolution in Closed Community.}
\newblock International Journal of Modern Physics C. 2000; 11(6): 1157.

\bibitem{javarone01}
Javarone MA.
\newblock{Social influences in opinion dynamics: the role of conformity.}
\newblock Physica A: Statistical Mechanics and its Applications. 2014; 414: 19--30.

\bibitem{javarone02}
Javarone MA.
\newblock {Networks strategies in election campaigns.}
\newblock Journal of Statistical Mechanics: Theory and Experiments. 2014; P08013.

\bibitem{redner01}
Sood V, Redner S.
\newblock{Voter Model on Heterogeneous Graphs.}
\newblock Phys. Rev. Lett. 2005; 94(17): 178701.

\bibitem{javarone04}
Javarone MA, Armano G.
\newblock{Emergence of Acronyms in a Community of Language Users.}
\newblock European Physical Journal - B. 2013; 86(11): 474.

\bibitem{perc05} 
D'Orsogna M, Perc M.
\newblock{Statistical physics of crime: A review.}
\newblock Phys. Life Rev. 2015; 12: 1--21.

\bibitem{galam03-bis}
Galam S.
\newblock{The September 11 attack: A percolation of individual passive support.}
\newblock European Physical Journal B. 2002; 26: 269--272.

\bibitem{galam03-ter}
Galam S.
\newblock{Global physics: from percolation to terrorism, guerilla warfare and clandestine activities.}
\newblock Physica A: Statistical Mechanics and its Applications. 2003; 330: 139--149.

\bibitem{javarone03}
Javarone MA, Galam S
\newblock{Emergence of extreme opinions in social networks.}
\newblock Lecture Notes on Computer Science, Springer. 2015.

\bibitem{moreno02}
Gracia-Lazaro C, Quijandria F, Hernandez L, Floria LM, Moreno Y.
\newblock{Co-evolutionary network approach to cultural dynamics controlled by intollerance.}
\newblock Phys. Rev. E. 2011; 84(6): 067101.

\bibitem{iglesias01}
Goncalves S, Laguna MF, Iglesias JR.
\newblock{Why, when, and how fast innovations are adopted.}
\newblock European Physical Journal - B. 2012; 85:192.

\bibitem{mcmillon01}
McMillon D, Simon CP, Morenoff J.
\newblock{Modeling the Underlying Dynamics of the Spread of Crime.}
\newblock PloS ONE. 2014; 9(4): e88923.

\bibitem{galam05}
Nizamani S, Memon N, Galam S
\newblock{From public outrage to the burst of public violence: An epidemic-like model.}
\newblock Physica A: Statistical Mechanics and its Applications. 2014; 416: 620--630.

\bibitem{galam04}
Qian S, Liu Y, Galam S.
\newblock{Activeness as a key to counter democratic balance.}
\newblock Physica A: Statistical Mechanics and its Applications. 2015; 432: 187--196.

\bibitem{galam03}
Galam S, Mauger A.
\newblock{On reducing terrorism power: a hint from physics.}
\newblock Physica A: Statistical Mechanics and its Applications. 2003; 323: 695--704.

\bibitem{islamic01}
\newblock{Network of terror: how DAESH uses adaptive social networks to spread its message.}
2015; Available:  \url{http://stratcomcoe.org/network-terror-how-daesh-uses-adaptive-social-networks-spread-its-message}.

\bibitem{holme01} 
Wu ZX, Holme P.
\newblock{Effects of strategy-migration direction and noise in the evolutionary spatial prisoner’s dilemma.}
\newblock Phys. Rev. E. 2010; 80(2): 026108.

\bibitem{perc03} 
Perc M, Grigolini P.
\newblock{Collective behavior and evolutionary games – An introduction.}
\newblock Chaos, Solitons \& Fractals. 2013; 56: 1-–5.

\bibitem{nowak01}
Nowak MA.
\newblock{Evolutionary Dynamics: Exploring the Equations of Life.}
\newblock Harvard University Press; 2006.

\bibitem{tomassini01} 
Tomassini M.
\newblock{Introduction to evolutionary game theory.}
\newblock Proc. Conf. on Genetic and evolutionary computation companion. 2014.

\bibitem{javarone05}
Javarone MA.
\newblock{Statistical Physics of the Spatial Prisoner's Dilemma with Memory-Aware Agents}
\newblock European Physical Journal - B. 2016; 89(42).

\bibitem{battiston01}
Battiston F, Cairoli A, Nicosia V, Baule A, Latora V.
\newblock{Interplay between consensus and coherence in a model of interacting opinions.}
\newblock Physica D \textit{In press}. 2016

\bibitem{li01}
Li Q, Braunstein LA, Wang H, Shao J, Stanley HE, Havlin S.
\newblock{Non-consensus opinion models on complex networks.}
\newblock Journal of Statistical Physics. 2013; 151: 92--112.

\bibitem{krokidakis01}
Crokidakis N, Castro de Oliveira PM.
\newblock{Inflexibility and independence: Phase transitions in the majority-rule model.}
\newblock Phys Rev E. 2015; 92: 062122.

\bibitem{krokidakis02}
Crokidakis N, Anteneodo C.
\newblock{Role of conviction in nonequilibrium models of opinion formation.}
\newblock Phys Rev E. 2012; 86: 061127.

\bibitem{krokidakis03}
Crokidakis N, Blanco VH, Anteneodo C.
\newblock{Impact of contrarians and intransigents in a kinetic model of opinion dynamics.}
\newblock Phys Rev E. 2014; 89: 013310.

\bibitem{pickering01}
Pickering W, Szymanski BK, Lim C.
\newblock{Opinion Diversity and the Stability of Social Systems: Implications from a Model of Social Influence;}
\newblock 2016. Preprint. Available: arXiv:1512.03390v3. Accessed 7 March 2016.

\bibitem{cheon01}
Cheon T, Morimoto J.
\newblock{Balancer effects in opinion dynamics.}
\newblock Physics Letters A. 2016; 380(3): 429-–434.

\bibitem{menezes01}
Oliveira M, Barbosa-Filho H, Yehle T, White S, Menezes R.
\newblock{From Criminal Spheres of Familiarity to Crime Networks.}
\newblock Studies in Computational Intelligence. 2015; 597: 219--230.

\bibitem{menezes02}
White S, Yehle T, Serrano H, Oliveira M, Menezes R.
\newblock{The Spatial Structure of Crime in Urban Environments.}
\newblock Lecture Notes in Computer Science. 2015; 8852: 102--111.

\bibitem{girvan01}
Burghardt K, Rand WM, Girvan M.
\newblock{Competing opinions and stubbornness: connecting models to data.}
\newblock SSRN; 2014.

\bibitem{xie01}
Xie J, Sreenivasan S, Korniss G, Zhang W, Lim C, Szymanski BK.
\newblock{Social consensus through the influence of committed minorities.}
\newblock Phys Rev E. 2011; 84: 011130.

\bibitem{diaz01}
Garcia-Diaz C, Zambrana-Cruz G, van Witteloostuijn A.
\newblock{Political spaces, dimensionality decline and party competition.}
\newblock Advances in Complex Systems. 2013; 16(6): 1350019.

\bibitem{dixit01}
Dixit AK, Weibull JW.
\newblock{Political polarization.}
\newblock Proceedings of the National Academy of Sciences. 2007; 104(18): 7351--7356.

\bibitem{galam06}
Galam S.
\newblock{Heterogeneous beliefs, segregation, and extremism in the making of public opinions.}
\newblock Phys Rev E. 2005; 71: 046123.

\bibitem{cioffi01}
Alizadeh M, Cioffi-Revilla C.
\newblock{Activation Regimes in Opinion Dynamics: Comparing Asynchronous Updating Schemes.}
\newblock Journal of Artificial Societies and Social Simulation. 2015; 18(3): 8.

\bibitem{javarone06}
Javarone MA, Squartini T.
\newblock{Conformism-driven phases of opinion formation on heterogeneous networks: the q-voter model case.}
\newblock Journal of Statistical Mechanics: Theory and Experiment. 2015; P10002.

\bibitem{galam02}
Galam S, Jacobs F.
\newblock{The role of inflexible minorities in the breaking of democratic opinion dynamics.}
\newblock Physica A: Statistical Mechanics and its Applications. 2007; 381: 366--376.

\bibitem{vespignani01}
Pastor-Satorras R, Vespignani A.
\newblock{Epidemic spreading in scale-free networks.}
\newblock Phys Rev Let. 2001; 86: 3200.

\bibitem{lagorio01}
Lagorio C, Migueles MV, Braunstein LA, Lopez E, Macri PA.
\newblock{Effects of epidemic threshold definition on disease spread statistics.}
\newblock Physica A: Statistical Mechanics and its Applications. 2009; 388: 755--763.

\bibitem{epidem01}
Bailey N.
\newblock{The Mathematical Theory of Infectious Diseases and its Applications.}
\newblock Griffin, London; 1975.

\bibitem{coles01}
Kelling GL, Coles CM.
\newblock{Fixing Broken Windows: Restoring Order and Reducing Crime in Our Communities.}
\newblock Simon and Schuster; 1997.

\bibitem{aronson01}
Aronson E, Wilson T, Akert RM.
\newblock{Social Psychology.}
\newblock Pearson Ed; 2006.
 

\end{thebibliography}
\end{document}